\documentclass[fleqn,twoside]{article}
\usepackage[headings]{espcrc2}
\usepackage{graphicx}
\newcommand{\MISR}[2]{{\cal M}^{\rm ISR (#2)}_{#1}}

\newcommand {\Sp}{{\rm Sp}\;}

\renewcommand\addressmark[1][\relax]{}%
\begin{document}
\title{Virtual Corrections to Bremsstrahlung in High-Energy Collider Physics:
	LHC and $e^+ e^-$ Colliders\thanks{Work supported in part by 
	U.S. Department of Energy contract DE-FG02-05ER41399.  S. Yost 
	thanks the organizers for an invitation to 
	present this work at RADCOR 2005.}}
\author{Scott A. Yost and B.F.L. Ward\address[Baylor]{Department of Physics, 
	Baylor University\\
	One Bear Place, P.O. Box 97316, Waco, TX 76798-7316}
       }
\runtitle{Virtual Corrections to Bremsstrahlung in High-Energy Collider Physics}
\runauthor{Scott A. Yost and B.F.L. Ward}
\begin{abstract}
We describe radiative corrections to bremsstrahlung and their application
to high energy collider physics, focusing on the applications to luminosity measurement, fermion pair production and radiative return. We review the
status of one loop radiative corrections in BHLUMI and the ${\cal KK}$MC, 
including cross checks with newer results developed independently for 
radiative return.  We outline a YFS-exponentiated approach to the Drell-Yan 
process for LHC physics, including a discussion of the relevant radiative 
corrections.
\vspace{1pc}
\end{abstract}
\maketitle
\thispagestyle{headings}
\renewcommand{\thepage}{\large $\hbox{\sf BU-HEPP-06/03}\atop \hbox{\hfill \sf January, 2006}$}

\section{RADIATIVE CORRECTIONS TO BHABHA SCATTERING}

The BHLUMI Monte Carlo (MC) program\cite{bhlumi} was developed as a precision 
tool for calculating the small-angle Bhabha luminosity process 
at SLC and LEP, and with continued development, it will continue to be a 
valuable tool meeting the requirements of a next-generation linear 
$e^+ e^-$ collider, such as the proposed ILC.  Central to this program's 
success was an exact treatment of the phase space for $n$ photon 
bremsstrahlung. A YFS-exponentiation\cite{yfs} procedure allows all IR 
singularities to be canceled exactly between real and virtual emission 
processes to all orders.  The leading soft photon effects are exponentiated, 
and IR-finite YFS residuals are then calculated exactly to the order 
required to reach the desired precision level.

BHLUMI attained a total error budget of 0.061\% for LEP1 parameters and 
0.122\% for LEP2 parameters for a typical calorimetric detector 
scenario.\cite{precision} To assure this precision level, it was 
necessary to calculate the most important unimplemented effect in BHLUMI4.04,
which was the next to leading-log (NLL) contribution to the two-photon
radiative corrections.  The double real photon and single real plus virtual 
photon corrections to the small-angle Bhabha scattering process were calculated
exactly in refs.\ \cite{2real,real+virt}  When added to the known two-loop 
virtual correction\cite{BVNB}, these results showed that the 
${\cal O}(\alpha^2 L)$ corrections enter at the 0.027\% level for LEP1 
parameters and 0.04\% level for LEP2 parameters. Here, $L = \ln(|t|/m_e^2)$ 
is the ``large logarithm'' entering into a leading log expansion.

Implementing these exact ${\cal O}(\alpha^2)$ results 
in BHLUMI would eliminate these contributions to the error budget.
The only remaining unimplemented ${\cal O}(\alpha^2)$ radiative corrections
would then be up-down interference effects in which two virtual photons
are exchanged between the $e^+$ and $e^-$ line, which are suppressed
at small angles, and nominally enter at the level of 0.004\% or less 
for angles below $9^\circ$.\cite{updown,snowmass} These contributions,
represented by diagrams of the type shown in Fig.\ 1,  could
thus be safely neglected for SLC and LEP physics. 

\begin{figure}[ht]
\kern-0.5cm\hbox{\kern-0.75cm\includegraphics*[width=9cm]{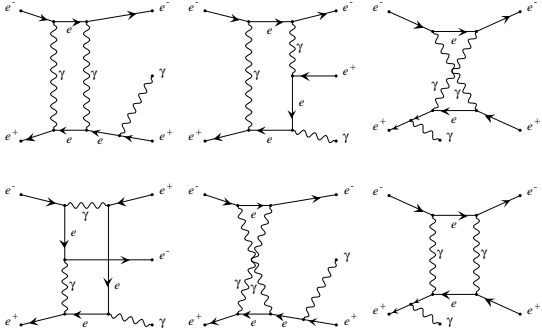}}
\vspace{-1.5cm}
\caption{One-loop up-down interference diagrams for $t$-channel positron 
line emission. The internal photons may also be replaced by a $Z$ boson.}
\end{figure}

For ILC physics, where the goal is to reach 0.01\% in the small-angle
Bhabha luminosity process, it is desirable to carefully check the magnitude 
of the up-down interference terms,
and to implement them if they turn out to be significant.
A key ingredient in the comparison, the five-point box integral appearing in
the second and fourth diagram in Fig.\ 1, has recently been provided by 
the Looptools 2.2 package.\cite{looptools}
A number of other ${\cal O}(\alpha^2)$ calculations which have appeared
recently\cite{penin,lorca} should also provide valuable insight into
effects which may need to be implemented in the small-angle Bhabha
calculation to reach ILC precision specifications.

\renewcommand{\thepage}{\arabic{page}}
\section{RADIATIVE CORRECTIONS TO FERMION PAIR PRODUCTION}

Fermion pair production plays a critical role in extracting precision
electroweak physics from $e^+ e^-$ colliders. This process is calculated
by another YFS-exponentiated MC program, the ${\cal KK}$MC.\cite{kkmc}
Again, the photonic radiative corrections play an essential role in 
calculating the YFS residuals through order $\alpha^2$. These have been
calculated exactly, including finite-mass corrections, for initial state
and final state radiation.\cite{jmwy} 

Using helicity spinor techniques,\cite{chinese-magic,KS} a concise 
and stable representation for the ${\cal O}(\alpha^2)$ initial or final state
radiation amplitude has been obtained, including finite-mass corrections.
 The matrix element for hard photon initial-state
emission with one virtual photon may be expressed as
\begin{equation}
\label{vdef1}
\MISR{1}{1} = {\alpha\over 4\pi} \MISR{1}{0} (f_0 + f_1 I_1 + f_2 I_2),
\end{equation}
where $\MISR{1}{0}$ is the tree-level matrix element for single hard photon
emission, 
$f_i$ are scalar form factors and $I_i$ are spinor factors defined in
ref.\ \cite{jmwy}.

The single hard photon cross section is of particular interest in radiative
return applications\cite{rad1,rad2,rad3}, where initial state radiation is 
used to reduce the effective beam energy, allowing a fixed energy machine to 
probe a range of energies. A MC program PHOKHARA was developed to calculate
radiative return at $\Phi$ and $B$ factories.\cite{rad4,PHOK2005} The same
radiative corrections are relevant for a high-energy $e^+ e^-$ collider
investigating physics around the $Z$ peak, for example. It is therefore useful 
to compare the radiative corrections obtained for both the ${\cal KK}$MC and 
PHOKHARA in detail. Both calculations claim the same level of exactness,
including the same diagrams as well as electron mass corrections relevant
for collinear bremsstrahlung.

We have compared the virtual corrections to initial state hard-photon 
emission calculated in ref.\ \cite{rad5,rad6} (KR) for PHOKHARA to those 
calculated in ref.\ \cite{jmwy} (JMWY) for the ${\cal KK}$MC in the case of 
muon pair production. Analytically, it was found that in the absence of 
mass corrections, both expressions agree to NLL order (${\cal O}(\alpha^2 L)$ 
in the integrated cross section).\cite{beijing} A compact
expression for NLL limit of the matrix element was obtained in 
ref.\ \cite{jmwy}, where it was shown that the terms $f_1$ and $f_2$ in
eqn.\ (\ref{vdef1}) vanish to NLL order, and the helicity-averaged NLL
limit of $f_0$ is 
\begin{eqnarray}
{\rm Re}\;\langle f_0^{\rm NLL}\rangle &=& 2\pi{\rm Re}\,B_{\rm YFS}(s) + L - 1
        \nonumber\\ 
&+& 2\ln r_1 \ln (1-r_2) - \ln^2(1-r_1) \nonumber\\
&+& 3\ln(1-r_1) + {r_1(1-r_1)\over 1 + (1-r_1)^2}  \nonumber\\
&+& 2\Sp(r_1) + (r_1 \rightarrow r_2).
\label{f0}
\end{eqnarray}
where $L =  \ln(s/m_e^2)$ is the ``large logarithm'' in the 
leading log expansion, $r_i = 2p_i\cdot k/s$ measures the inner product of
one of the incoming fermion momenta $p_i$ with the hard photon momentum $k$,
Sp$(x)$ = Li$_2(x)$ is the dilogarithm (Spence) function, and 
\begin{eqnarray}
4\pi {\rm Re}\,B_{\rm YFS}(s) &=& \left(2\ln{m_\gamma^2\over m_e^2} + 1\right)
        \left(L - 1\right) \nonumber\\
&-& L^2 - 1 + {4\pi^2\over3}
\end{eqnarray}
is the IR-divergent virtual YFS form factor. 

Since the two results are known
to agree with the NLL limit calculated using eqn.\ (\ref{f0}), the NLL limit
is subtracted in each case, permitting the NNLL contributions and collinear
mass corrections to be investigated in the context of the ${\cal KK}$MC.
Fig.\ 2 shows the results of a ${\cal KK}$MC run calculating the NNLL 
contribution to muon pair production at a CMS energy of 500 GeV
for $10^8$ events, both with and without the mass corrections. The cross-section
is integrated up to a radiated photon energy fraction of $v_{\rm max}$
(with $v = r_1 + r_2$) using the YFS residual $\beta_1^{(2)}$ for one hard 
photon at ${\cal O}(\alpha^2)$, subtracting the NLL contribution obtained using
eqn.\ (\ref{f0}). The result is normalized with respect to the non-radiative
Born cross section for muon pair production. 

\begin{figure}[ht]
\includegraphics*[width=7.5cm]{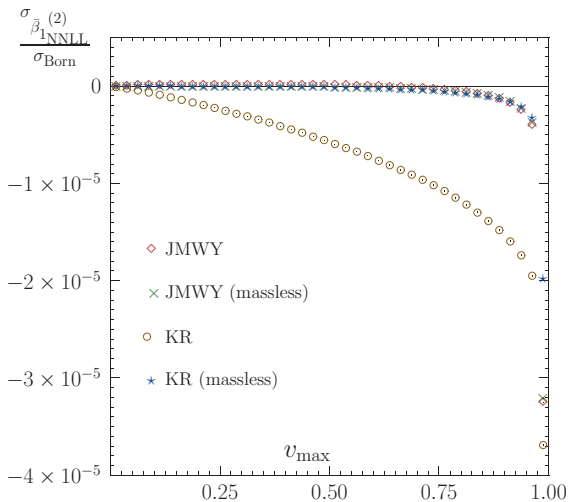}
\vspace{-1cm}
\caption{Comparison of NNLL contributions to the virtual correction to the 
single hard photon cross section for muon pair production at a CMS energy
of 500 GeV. }
\end{figure}

It is found that the maximum
difference between the complete KR and JMWY results (from the next-to-last
bin) is $1.6\times 10^{-5}$ units of the Born cross section. Most of this
apparently comes from differences in the treatment of the mass corrections.
KR uses an expansion in $m_e^2/r_is$, while JMWY uses a technique developed
by Berends {\it et al.}\cite{berends1} for adding the mass corrections 
required in collinear limits to a calculation obtained using massless spinors.
Without mass corrections (comparing the massless points), the results 
agree to within a part per million. This agreement is better than noted 
previously\cite{beijing,paris,compare,epiphany} due to improvements in the 
stability of the algorithms used.  Direct comparisons of the PHOKHARA and 
${\cal KK}$MC programs have also been conducted.\cite{hans2,jadcomp}

It is interesting to note that the size of the NNLL part of the corrections
implemented by JMWY never exceed $4\times 10^{-6}$ up to $v_{\rm max} = 0.975$,
and reach $-3.25\times 10^{-5}$ in the last bin, where $v_{\rm max} = 0.9875$.
This suggests that for most purposes, the considerably simpler NLL result
represented by eqn.\ (\ref{f0}) will suffice.

\section{THE DRELL-YAN PROCESS}

The Drell-Yan process plays a role at hadron colliders which is as basic as
the Bhabha scattering or pair production cross section at $e^+ e^-$ colliders.
In fact, $W$ and $Z$ production has been proposed as the luminosity process
for the LHC.\cite{lumLHC}  A fully exclusive calculation of the parton-level 
cross sections is needed at the $1-2\%$ level for upcoming LHC physics. 
These cross sections are currently known at the $10\%$ level, using NLO 
matrix elements.  

While NNLO results are available for the integrated cross 
section\cite{drell-yan} and rapidity distribution\cite{rapidity}, 
a fully-exclusive NNLO cross section needed for
a MC event generator is not yet available.  Moreover, electroweak radiative
corrections will be required as well. Reaching
the desired LHC precision will require corrections of order $\alpha_{\rm s}^2$ 
and order $\alpha_{\rm ew}$, including mixed 
${\cal O}(\alpha_{\rm s}\alpha_{\rm ew})$ contributions. Examples of the 
latter diagrams are shown in Fig.\ 3.

\begin{figure}[ht]
\setlength{\unitlength}{1cm}
\begin{picture}(7.5,6.5)
\put(1.8,3.5){(a)}
\put(0,3.25){\hbox{\includegraphics*[width=4cm]{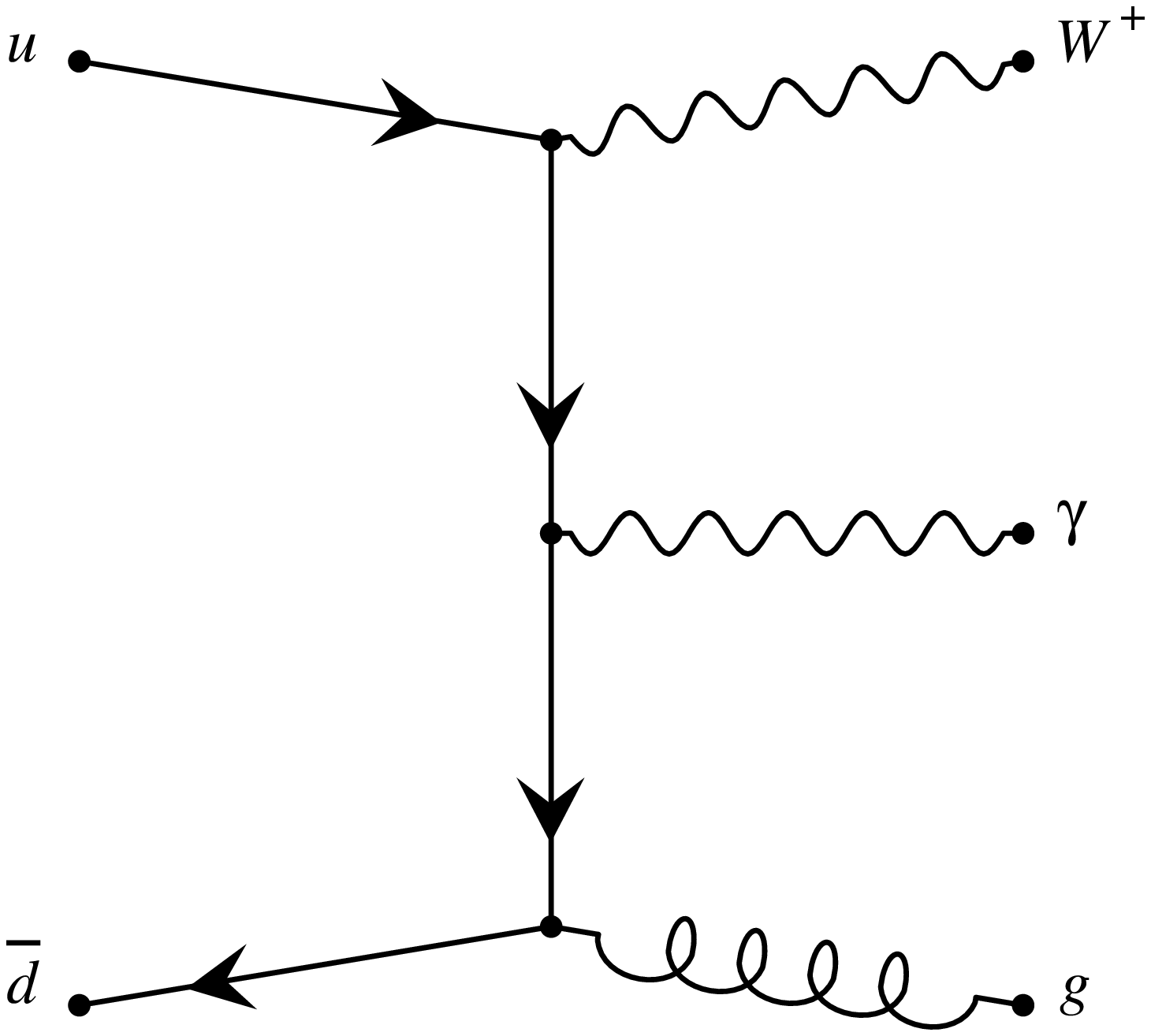}}}
\put(5.6,3.5){(b)}
\put(0,0){\hbox{\includegraphics*[width=4cm]{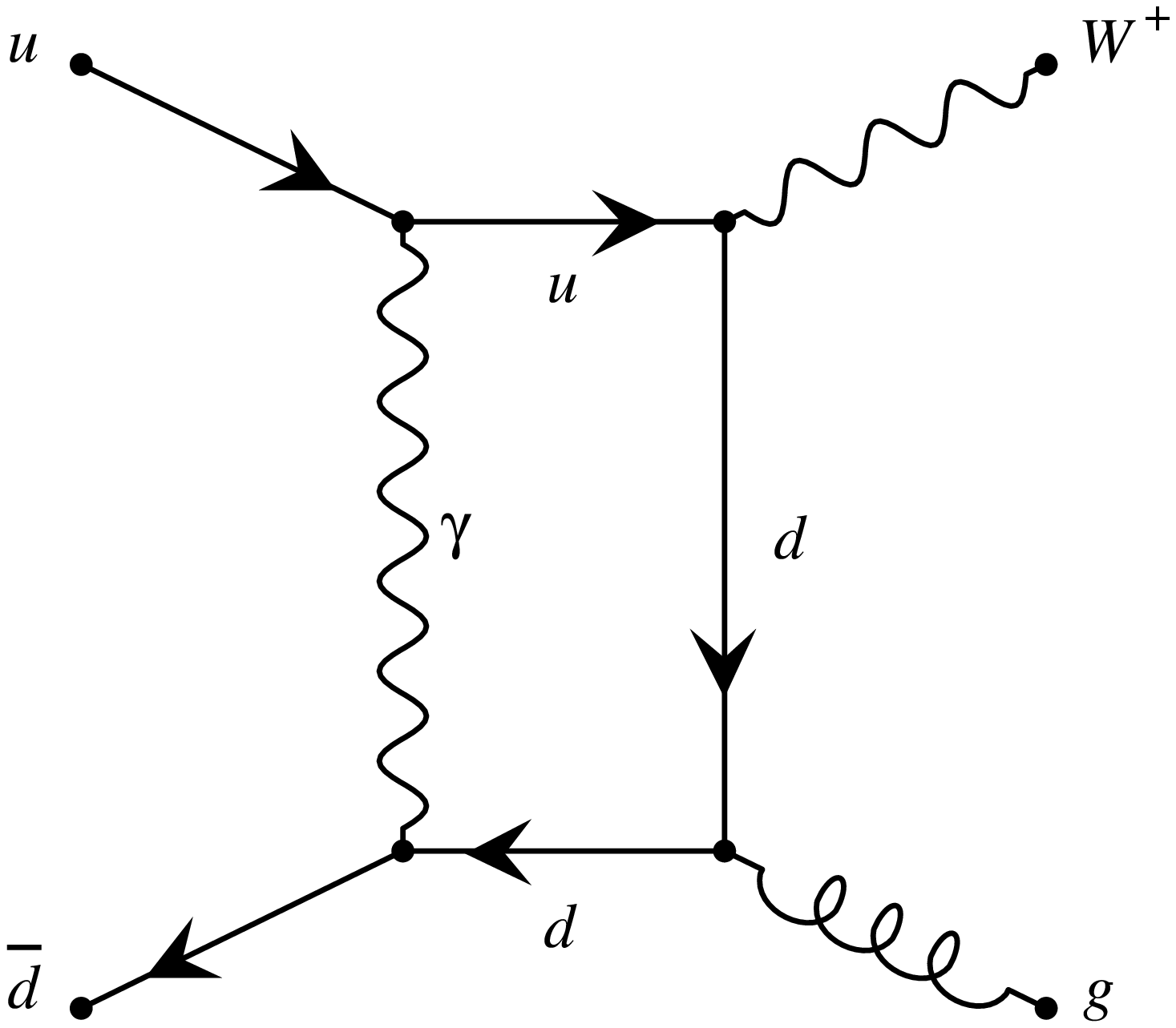}}}
\put(1.8,0.25){(c)}
\put(3.75,3.25){\hbox{\includegraphics*[width=4cm]{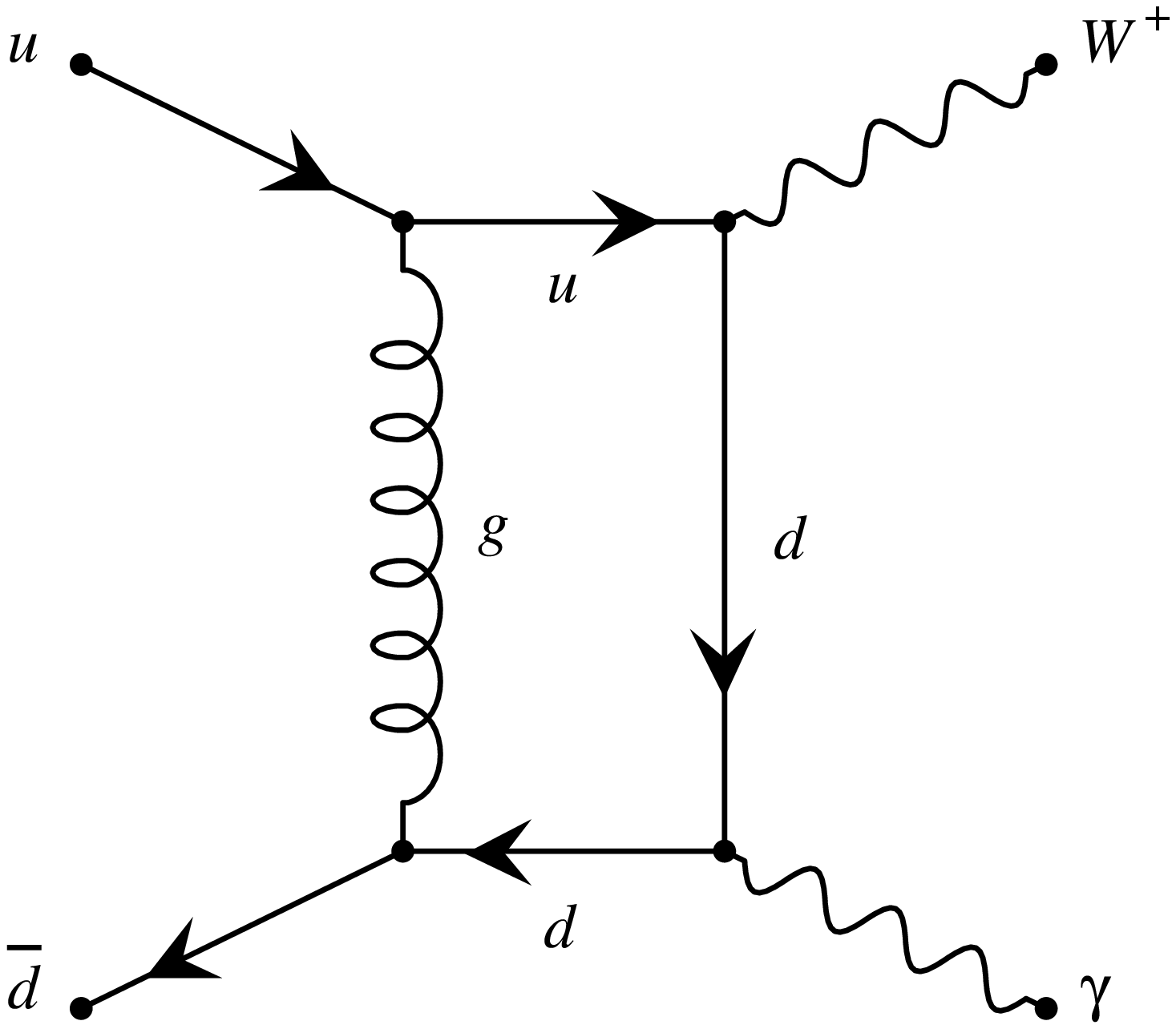}}}
\put(5.6,0.25){(d)}
\put(3.75,0){\hbox{\includegraphics*[width=4cm]{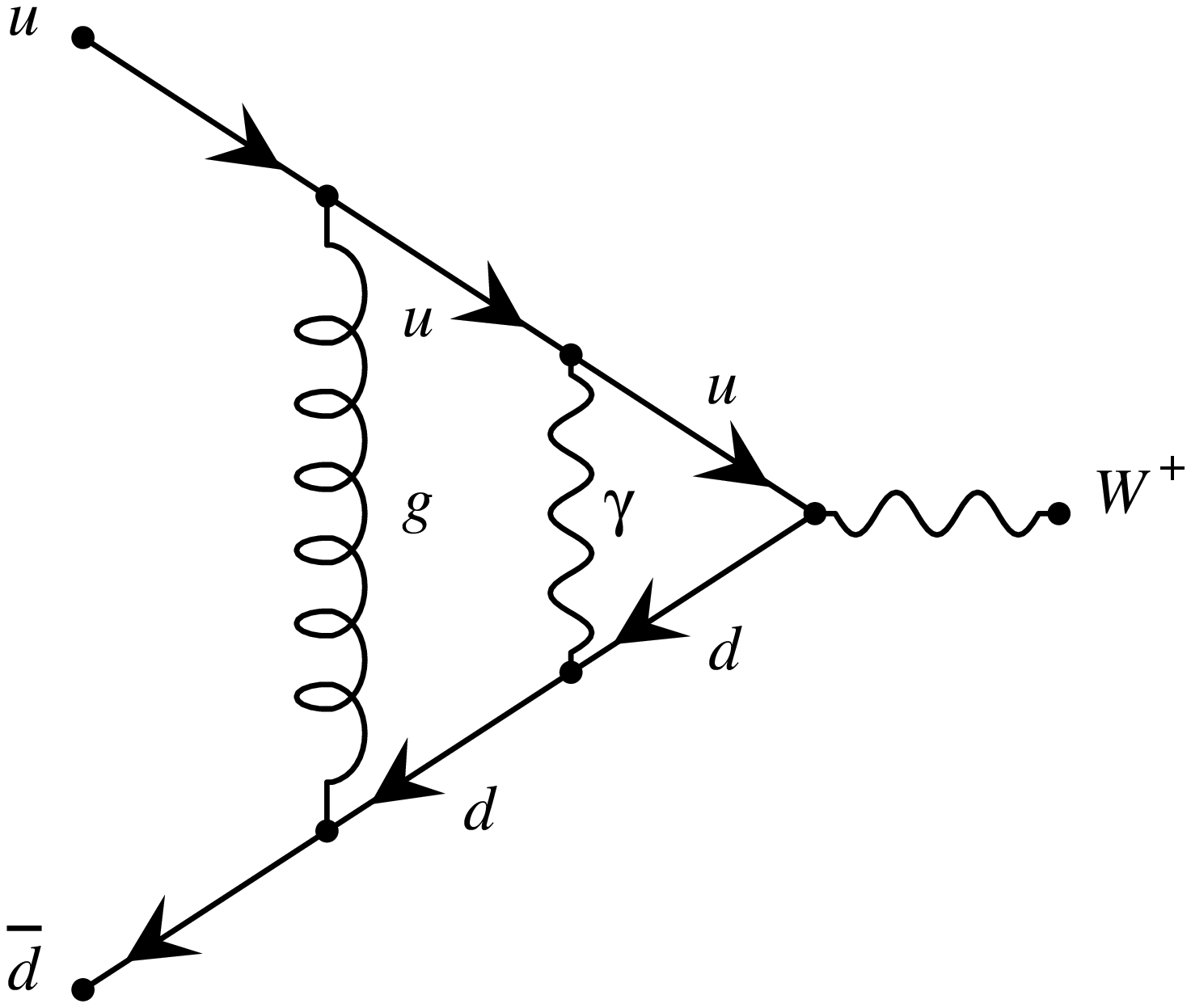}}}
\end{picture}
\vspace{-1.25cm}
\caption{Examples of mixed photonic and gluonic initial-state radiative 
corrections: (a) real photon + gluon emission, (b) real photon with QCD loop, 
(c) real gluon with electroweak loop, (d) 2 loops, mixed QCD/electroweak.
The final state fermions from the $W^+$ decay are not shown.}
\end{figure}

The hard parton-level processes must be calculated and combined with PDFs
in a MC program designed to generate the desired distribution of partons
plus mixed QCD and QED bremsstrahlung.  This will require a careful 
implementation of the multiple gluon and photon phase space. Experience in the 
electroweak sector suggests that YFS exponentiation will provide a strong tool
for implementing the multiple-emission phase space, giving very precise
control over the soft and collinear limits. 

The large numbers of diagrams creates a challenge for obtaining an expression
that can be evaluated quickly enough for MC implementation, and evaluated
in a numerically stable manner. Common reduction methods based on the
Passarino-Veltman technique\cite{PV} can produce millions of terms, which are
both slow to evaluate in a MC setting, and prone to numerical instabilities
due to the large numbers of terms added and potential cancellations among them.
Thus, a significant part of this problem will involve developing and testing
new methods for organizing and calculating the terms in a stable manner.

Once the parton-level matrix element is obtained, it may be incorporated
into a ``QCED-Exponentiated'' MC program, implementing a procedure similar
to YFS exponentiation in a combined QED and QCD setting to construct the
exact phase space for multiple gluon and photon radiation. This requires
extending the YFS calculus to non-abelian gauge theory, with due care
in handling the genuine non-abelian singularities which arise in QCD.
The QCD and QED exponentiation will be conducted at order $\alpha_{\rm s}^2 L$
on an event-by-event basis in the presence of showers without double-counting
of soft and collinear emission effects. Further details of this construction
will be presented elsewhere in these proceedings.\cite{QCED-RADCOR}

\section{SUMMARY AND OUTLOOK}

A careful calculation of higher-order brems\-strahlung corrections led to 
a precision tool (BHLUMI) for Bhabha luminosity calculations. 
Incorporating the second order photonic corrections obtained to test 
BHLUMI's precision may be enough to reach the 0.01\% level proposed for the 
ILC.  Calculating the ${\cal O}(\alpha^2)$ up-down interference contribution
will help to clarify this. A number of 
second-order Bhabha scattering results are now available which should be
useful for testing BHLUMI's precision at this higher level. 

We have also described cross-checks on the second-order photonic radiative 
corrections for fermion pair production developed for the ${\cal KK}$MC.
Comparisons with similar initial-state radiative corrections developed
for PHOKHARA show agreement on the order of $10^{-5}$ in units of the 
Born cross section at ILC energies.  Remaining differences may
be attributed primarily to differences in the handling of finite mass
corrections in the collinear limits, and possible residual numerical
instabilities for photons radiated near the fermion pair production threshold. 

Finally, we have outlined a program for carrying the successes of the 
YFS-exponentiated MC framework developed for SLC and LEP physics into the 
hadronic realm of the LHC, where a precise calculation of the parton-level
diagrams can be combined with a combined QCD and QED exponentiation framework,
with complete control of the multiple photon and gluon phase space,
to develop a MC for the Drell-Yan process which can reach the LHC precision
requirements.


\begin{thebibliography}{99}
                                                                                
\bibitem{bhlumi}
S.\ Jadach, E. Richter-W{\c a}s, W. P{\l}aczek, B.F.L.\ Ward and Z. W{\c a}s,
Comp.\ Phys.\ Commun.\ 102 (1997) 229.
                                                                                
\bibitem{yfs} D.R.\ Yennie, S.C.\ Frautschi, and H.\ Suura, {Ann.\ Phys.}
{13} (1961) 379; K.T.\ Mahanthappa, {Phys.\ Rev.} {126} (1962) 329.

\bibitem{precision}
B.F.L.\ Ward, S. Jadach, M. Melles and S.A.\ Yost, {Phys.\ Lett.}
{B450} (1999) 262.
                                                                                
\bibitem{2real}
S.\ Jadach, B.F.L.\ Ward and S.A.\ Yost, Phys.\ Rev.\ D47 (1993) 2682.
                                                                                
\bibitem{real+virt}
S.\ Jadach, M.\ Melles, B.F.L.\ Ward and S.A.\ Yost, Phys.\ Lett.\ B377 (1996)
168.

\bibitem{BVNB}
F.A.\ Berends, W.L.\ Van Neerven and G.J.H.\ Burgers, Nucl.\ Phys.\ B297 (1988)
429.

\bibitem{updown}
S. Jadach, E. Richter-W{\c a}s, B.F.L.\ Ward and Z. W{\c a}s, Phys.\ Lett.\ 
B253 (1991) 469.

\bibitem{snowmass}
S.A.\ Yost, S. Majhi and B.F.L.\ Ward, Proceedings of the 2005 International 
Linear Collider Physics and Detector Workshop and Second ILC Accelerator
Workshop, Snowmass (2005), ALCPG1911 (hep-ph/0512022).


\bibitem{looptools}
T. Hahn, Comput.\ Phys.\ Commun.\ 118 (1999) 153; T. Hahn, proceedings of 
RADCOR 2005.

\bibitem{penin} A.A.\ Penin, Phys.\ Rev.\ Lett.\ 95 (2005) 010408;
Nucl.\ Phys.\ B734 (2006) 185; proceedings of RADCOR 2005.
                                                                                
\bibitem{lorca} A. Lorca and T. Reimann, DESY report 04-226 (hep-ph/0412047); 
A. Lorca, contribution to ACAT 2005 (hep-ph/0509367).

\bibitem{kkmc}
S.\ Jadach, B.F.L.\ Ward and Z.\ W{\,}as, {Phys.\ Rev.} {D63} (2001)
113009; {Comput.\ Phys.\ Commun.} {130} (2000) 260.
                                                                                
\bibitem{jmwy}
S. Jadach, M. Melles, B.F.L.\ Ward and S.A.\ Yost, {Phys.\ Rev.} {D65}
 (2002) 073030.

\bibitem{chinese-magic}
Z.\ Xu, D.-H.\ Zhang and L.\ Chang, {Nucl.\ Phys.} {B291}
(1987) 392.
                                                                                
\bibitem{KS}
R. Kleiss and W.J.\ Stirling, {Nucl.\ Phys.} {B262} (1985) 235;
Phys.\ Lett.\ B{179} (1986) 159.
                                                                                
\bibitem{rad1}
S. Binner, J.H.\ K\"uhn, K. Melnikov, {Phys.\ Lett.}\ {B459} (1999) 279.
                                                                                
\bibitem{rad2}
K. Melnikov, F. Nguyen, B. Valeriani, G. Venanzoni,
{Phys.\ Lett.}\ {B477} (2000) 114.
                                                                                
\bibitem{rad3}
H. Czy\.z, J.H. K\"uhn, {Eur.\ Phys.\ J.}\ {C18} (2001) 497.
                                                                                
\bibitem{rad4}
G. Rodrigo, H. Czy\.z, J.H. K\"uhn, M. Szopa, {Eur.\ Phys.\ J.}\ {C24}
(2002) 71.
                                                                                
\bibitem{PHOK2005}
H. Czy\.z, A. Grzeli{\'n}ska and E. Nowak-Kubat, Acta Phys.\ Polon.\ B36
(2005) 3403.

\bibitem{rad5}
G. Rodrigo, A. Gehrmann-De Ridder, M. Guilesaume, J.H.\ K\"uhn, Eur.\ Phys.\ J.
 {C22} (2001) 81.
                                                                                
\bibitem{rad6}
J.H.\ K\"uhn, G. Rodrigo, {Eur.\ Phys.\ J.}\ {C25} (2002) 215.
                                                                                
\bibitem{beijing}
S.A.\ Yost, C. Glosser, S. Jadach and B.F.L.\ Ward, ICHEP 2004: Proceedings of
the 32$^{\rm nd}$ International Conference on High Energy Physics, Beijing
(World Scientific, Singapore, 2005) 478.

\bibitem{berends1}
F.A.\ Berends, R. Kleiss, P. De Causmaecker, R.
Gastmans, W. Troost, T.T.\ Wu, {Nucl.\ Phys.}\ {B206}, (1982) 61.

\bibitem{paris}
S.A.\ Yost, C. Glosser, S. Jadach and B.F.L.\ Ward, Proceedings of the 
International Conference on Linear Colliders, LCWS04, Paris (2004)
(hep-ph/0404087).

\bibitem{compare}
C. Glosser, S. Jadach, B.F.L.\ Ward and S.A.\ Yost, Phys.\ Lett.\ B605
(2005) 123.

\bibitem{epiphany}
S.A.\ Yost, S. Jadach and B.F.L.\ Ward, Acta Phys.\ Polon.\ B36 (2005) 2379.

\bibitem{hans2}
H. Czy\.z, A. Grzeli{\'n}ska, J.H.\ K\"uhn and  G. Rodrigo, {SIGHAD03:
Workshop on Hadronic Cross Section at Low Energy, Pisa} (2003)
\hbox{(hep-ph/0312217).}
                                                                                
\bibitem{jadcomp}
S. Jadach, {Acta Phys.\ Polon.}\ {B36} (2005) 2387.
                                                                                
\bibitem{lumLHC}
M. Dittmar, F. Pauss and D. Z\"urcher, Phys.\ Rev.\ D56 (1997) 7284;
V.A.\ Khoze, A.D.\ Martin, R. Orava and M.G.\ Ryskin, Eur.\ Phys.\ J.\ C19 
(2001) 313.

\bibitem{drell-yan}
R. Hamberg, W.L.\ van Neerven and T. Matsuura, Nucl.\ Phys.\ B359 (1991) 343;
R.V.\ Harlander and W.B.\ Kilgore, Phys.\ Rev.\ Lett.\ 88 (2002) 201801.

\bibitem{rapidity}
C. Anastasiou, L. Dixon, K. Melnikov, and F. Petriello, Phys.\ Rev.\ Lett.\ 91
(2003) 182002, Phys.\ Rev.\ D69 (2004) 094008.

\bibitem{QCED-RADCOR}
B.F.L.\ Ward, proceedings of RADCOR 2005.

\bibitem{PV}
G. Passarino and M. Veltman, Nucl.\ Phys.\ B160 (1979) 151.

\end{thebibliography}
\end{document}